\documentclass[prb,aps,twocolumn,showpacs]{revtex4}
\usepackage{amsmath,amssymb,amsfonts,float}
\usepackage[dvips]{graphicx,color}    
\newcommand{\bolT}{\text{\bf T}}
\newcommand{\bolK}{\text{\bf K}}

\newcommand{\bolk}{\mathbf{k}}
\newcommand{\bolp}{\text{\bf p}}

\newcommand{\bolq}{\mathbf{q}}

\newcommand{\bolQ}{\mathbf{Q}}

\newcommand{\bolr}{\mathbf{r}}
\newcommand{\boll}{\mathbf{l}}

\newcommand{\ket}[1]{| #1 \rangle}  
\newcommand{\VEV}[1]{\langle #1 \rangle}

\newcommand{\be}{\text{e}}

\newsavebox{\dotdot}
\savebox{\dotdot}[3mm]{\shortstack{\circle*{0.8}\\ \\ \circle*{0.8}}}

\begin{document}
\title{Nematic phase and phase separation near saturation field in frustrated ferromagnets
}
\author{Hiroaki~T.~Ueda$^{1}$ and Tsutomu~Momoi$^2$}
\affiliation{
$^1$Department of Physics, Tokyo Metropolitan University, Hachioji, Tokyo 192-0397, Japan\\
$^2$Condensed Matter Theory Laboratory, RIKEN, Wako Saitama 351-0198, Japan}
\begin{abstract}
We study the effects of quantum fluctuations on the magnetic properties of
quantum frustrated ferromagnets in a magnetic field.
It is shown that
a non classical phase or a phase separation appears due to quantum fluctuations
below the saturation field
in a parameter range close to the classical zero-field phase boundary
between ferromagnetic
and antiferromagnetic phases,
for the case that the classical antiferromagnetic state is not an eigenstate of the quantum model.
As an example to which this argument is applicable,
we study the $S=1/2$ $J_1$-$J_2$ Heisenberg model with ferromagnetic $J_1$
($J_1<0$) on the bcc lattice using a dilute Bose gas approach.
For $-1.50097 \leq J_1/J_2 \leq -1.389$, magnons form $f$-wave two-magnon bound states,
leading to a spin nematic phase, and for $-1.389 \leq J_1/J_2 \leq -0.48$, a canted
coplanar antiferromagnetic phase appears accompanied with a phase separation below the saturation field.
\end{abstract}
\maketitle
\section{introduction}
Magnetic frustration in
spin systems induces energy competition between different magnetic phases.
In a strongly competing regime,
quantum fluctuations are expected to
lift the degeneracy inducing a novel quantum phase different from the competing classical phases.
It has been revealed that quantum frustrated ferromagnets, in which ferromagnetic (FM)
and antiferromagnetic (AF) spin-exchange interactions strongly compete,
exhibit various exotic quantum phases.\cite{Chubukov,Vekua,Kecke,Hikihara,Sudan%
,SquareNem,NemTriangle1,OctTriangle,NemTriangle2}
Much attention has recently been paid to one of the
quantum phases, a spin nematic phase, which does not have any spin vector order
but exhibits a long range order of rank-2 spin tensors.

Recent theoretical studies revealed that spin nematic phases
exist in $S=1/2$ $J_1$-$J_2$ Heisenberg models with ferromagnetic $J_1$ ($J_1<0$)
and competing antiferromagnetic $J_2$ ($J_2>0$)
on the one-dimensional zigzag chain and on
the two-dimensional square lattice.
The Hamiltonian for the $J_1$-$J_2$ Heisenberg model is given by
\begin{equation}
{\cal H}=J_1 \sum_{\langle i,j \rangle} {\bf S}_i\cdot {\bf S}_j
+J_2 \sum_{\langle\langle i,j \rangle\rangle} {\bf S}_i\cdot {\bf S}_j
+ {\rm H}\sum S^z_i,
\label{HSpin}
\end{equation}
where $J_1$ ($J_2$) denotes the ferromagnetic nearest-neighbor
(antiferromagnetic next-nearest-neighbor) coupling and $H$ is an applied magnetic field.
The appearance of
a spin nematic phase is most established in the $S=1/2$ $J_1$-$J_2$ zigzag chain.
It is theoretically shown that the spin nematic phase is stabilized
under high magnetic field in the range $-2.72<J_1/J_2<0$.\cite{Chubukov,Vekua,Hikihara}
These theoretical works motivated recent active experimental
searches\cite{NemExp,NemExp_2,NemExp_3} for spin nematic phases
in the quasi-one-dimensional $J_1$-$J_2$ compound\cite{LiCuVO4_1,LiCuVO4_2} LiCuVO$_4$.
On the square lattice,
both the numerical and analytical approaches\cite{SquareNem} showed
the existence of the spin nematic phase around the classical phase boundary ($J_1/J_2= -2$)
between ferromagnetic and collinear antiferromagnetic
phases;
below the saturation field,
the spin-nematic phase is firmly induced by the two-magnon
instability for the broad parameter range $-2.5 \lesssim J_1/J_2\lesssim -0.2$,\cite{footnote}
though there is still debate
about the stability of the spin-nematic order at zero
field.\cite{Richtersquare,Feldner,sm,ShindouM}
There are various compounds, e.g., BaCdVO(PO$_4$)$_2$, which are adequate
to consider as the square-lattice ferromagnetic $J_1$-$J_2$ model.\cite{J1J2compound_1,J1J2compound_2}
It is hence expected that the spin nematic phase
appears in these compounds, at least in a high-magnetic-field regime.\cite{footnote}
In addition to $J_1$-$J_2$ models, a spin nematic phase is also found in a
frustrated multiple-spin-exchange model on the triangular lattice,
which describes the magnetic properties of thin films of solid $^3$He.\cite{NemTriangle2}
This spin nematic phase also appears in a parameter range close to the classical phase boundary
between ferromagnetic and antiferromagnetic phases.

In this paper, we study the effects of quantum fluctuations in frustrated ferromagnets,
especially in the parameter range surrounding the classical boundary between ferromagnetic and
antiferromagnetic phases,
exploring microscopic spin models for spin nematic phases.
In Sec.~\ref{Sec:Gen}, we analyze the quantum fluctuations
at the classical boundary between ferromagnetic
and antiferromagnetic phases in general spin models. This argument concludes
that, if the classical antiferromagnetic state is not an eigenstate of the quantum Hamiltonian,
there must appear quantum phenomena below the saturation field which cannot be
described with the one-magnon instability
at the saturation. This implies that a simple canted antiferromagnetic phase is
not a stable state of matter below the saturation in the quantum frustrated magnets close to
the ferromagnetic phase boundary.
This general argument would
be useful to find quantum phenomena such as spin nematic order and phase separation
in frustrated magnets.

%
%

As an example to which this argument is applicable, we study
the magnetic structure slightly below the saturation field
in the three-dimensional $S=1/2$ $J_1$-$J_2$ model on the bcc lattice,
in Secs.~III and \ref{sec:BMBEC}.
This model has been extensively studied for antiferromagnetic couplings $J_1$ and $J_2$
as one of the minimal frustrated
magnets.\cite{Shender,bccHeisenberg_1,bccHeisenberg_2,bccHeisenberg_3}
Recent theoretical studies concluded that in the antiferromagnetic case ($J_{1,2}>0$)
the classically expected antiferromagnetic orders persist even in a quantum case.\cite{bccHeisenberg_1,bccHeisenberg_2,bccHeisenberg_3}
However, for the ferromagnetic coupling $J_1<0$, the argument in Sec.~\ref{Sec:Gen}
assures the appearance of a non-classical behavior in a magnetization process.
For a quantitative study, we adopt the dilute-Bose gas approach from
the viewpoint of the magnon Bose-Einstein condensation (BEC);
the emergent magnetic order below the saturation field can be viewed as a condensation of magnons.
This approach has succeeded to explain various experimental results.\cite{Giamarchi-R-T-08,Giamarchi-R-T-08_1,Nikuni-O-O-T-00,Radu}
Applying this approach, we found the appearance of a spin nematic phase
or a phase separation under high-magnetic field in the $J_1$-$J_2$ model on the bcc lattice.
Section~VII is devoted to the conclusion.

\section{General analysis}
\label{Sec:Gen}
In this section, we first analyze a tendency for the ferromagnetic phase boundary
to shift which commonly occurs in quantum frustrated ferromagnets.
Magnetism in frustrated ferromagnets depends
on the energy balance in exchange couplings:
strong ferromagnetic couplings stabilize a fully polarized ferromagnetic phase and
strong antiferromagnetic couplings induce an antiferromagnetic ordered phase.
In the classical system, the transition between the ferromagnetic phase and
the antiferromagnetic phase is usually first order.
In a quantum case, quantum fluctuations can give a competing regime
room to induce new quantum phenomena.

Let us start with an analysis of quantum fluctuations
at the classical boundary between ferromagnetic and antiferromagnetic phases in zero field.
%
%
In most of quantum systems, the classical
antiferromagnetic state $\ket{\text{AF;CL}}$
is not an eigenstate of the Heisenberg Hamiltonian,
${\cal H}\ket{\text{AF;CL}}=E_{\text{CL}}\ket{\text{AF;CL}}+\ket{\alpha}$,
where $E_{\text{CL}}$ denotes the ground-state energy of the classical antiferromagnetic state
and $\ket{\alpha}$ satisfies $\langle\text{AF;CL}\ket{\alpha}=0$ and $\langle \alpha \ket{\alpha}> 0$,
albeit
the fully polarized ferromagnetic state $\ket{\text{FM;CL}}$ is one of the eigenstates
having the same energy $E_{\text{CL}}$ at the classical boundary,
${\cal H}\ket{\text{FM;CL}}=E_{\text{CL}}\ket{\text{FM;CL}}$.
%
%
In this case, the variational principle guarantees that
the true quantum ground state at the classical FM/AF phase boundary
has an energy lower than that of the ferromagnetic state, $E_{\text{CL}}$.
Therefore, the zero-field boundary of the ferromagnetic phase in the quantum model shifts backward
from the classical boundary into the classical ferromagnetic phase region.
%

\begin{figure}[tb]
\begin{center}
\includegraphics[scale=0.25]{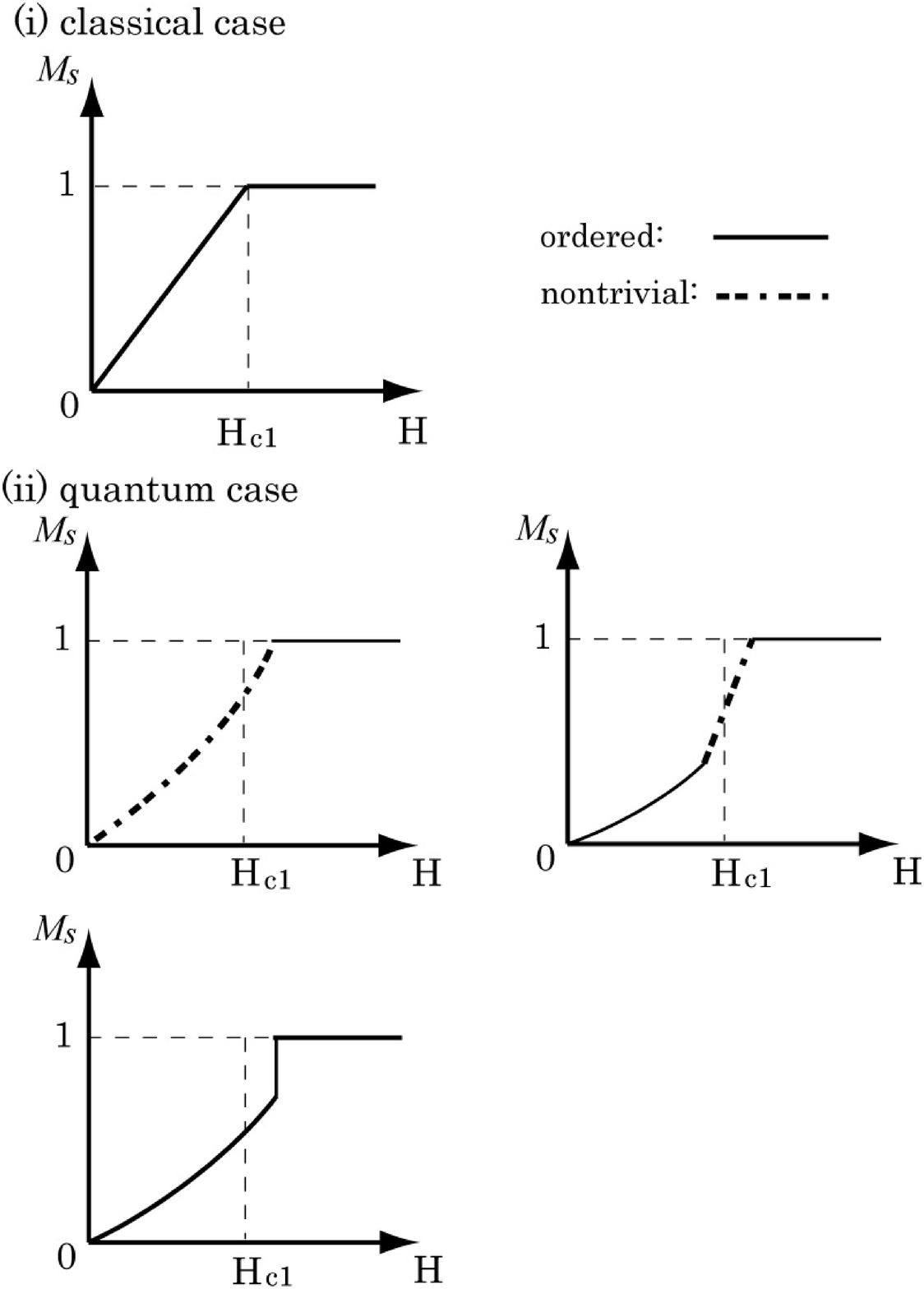}
\caption{Schematic behaviors of magnetization curves in frustrated ferromagnets
near the ferromagnetic/antiferromagnetic phase boundary.
The magnetization $M_s$ is given by $-(1/NS)\sum_{i=1}^N\VEV{S_i^z}$
and $\text{H}_{c1}$ denotes the saturation field in the classical limit.
The solid lines represent the antiferromagnetic spin-ordered phases and
the dashed lines represent nontrivial quantum phases.
If quantum fluctuation is taken into account, a nontrivial quantum phase or
a phase separation must appear near the saturation field.
\label{Fig;SCmag}}
\end{center}
\end{figure}

The magnetization process in applied field must be compatible with this boundary shift.
In the classical case,
the antiferromagnetic phase in a magnetic field is given by canted antiferromagnetic states
and this phase terminates at the saturation field ${\rm H}=\text{H}_{c1}$.
Even in quantum cases, the saturation field defined by one-magnon flips is the same
as the classical value $\text{H}_{c1}$ and
the condensation of single magnons below this saturation field leads to
the canted antiferromagnetic state.
As in the classical case, this saturation field $\text{H}_{c1}$ must vanish at
the classical FM/AF phase boundary.
%
In quantum systems, however, the true zero-field ground state
is not the ferromagnetic state as discussed above and hence
the true saturation field, which remains finite at the classical zero-field boundary,
vanishes at the quantum boundary to the ferromagnetic phase.
This concludes that the true saturation field is {\it not} given
by the single-magnon instability which induces the canted antiferromagnetic ordered phase;
the canted antiferromagnetic phase described with the single-magnon BEC\cite{footnote2}
below the saturation field
is veiled, in the vicinity of the classical FM/AF phase boundary
by emergence of a \emph{new quantum phase} or a \emph{phase separation},
as shown in Fig.~\ref{Fig;SCmag}.
An appealing alternative to the single-magnon BEC in applied field
is a BEC of bound multiple magnons.\cite{SquareNem,NemTriangle1,OctTriangle,NemTriangle2}

This argument can be applied to the phase boundary between the ferromagnetic
and collinear antiferromagnetic phases
in the square-lattice $J_1$-$J_2$ model\cite{SquareJ1J2} and also
to the boundary between the ferromagnetic and non-trivially degenerate paramagnetic phases
in the multiple-spin-exchange model on the triangular
lattice.\cite{TriMSEMF,TriMSEMC}
In both cases, the classical antiferromagnetic states are not eigenstates of 
the quantum Hamiltonian at the
classical phase boundary and indeed the appearance of spin
nematic\cite{NemTriangle1,SquareNem,NemTriangle2} and spin
triatic (octupolar)\cite{OctTriangle}
phases, originated from bound-multiple-magnon BEC, was theoretically proposed
in the vicinity of the classical phase boundary.
Naturally, this general analysis is applicable to $J_1$-$J_2$ models on different lattices,
even in three dimensions, where the quantum fluctuation is believed to be weak.
For example, the $J_1$-$J_2$ models on the bcc, the fcc, and the cubic
lattices satisfy the condition of this theorem.
Regarding the fcc $J_1$-$J_2$ model, a recent numerical study
by the exact diagonalization method found a phase separation in the magnetization
process in a parameter range near the classical ferromagnetic and
collinear antiferromagnetic phase boundary.\cite{FCCJ1J2}

\section{classical phases in the bcc lattice}
\begin{figure}[tb]
\begin{center}
\includegraphics[scale=0.3]{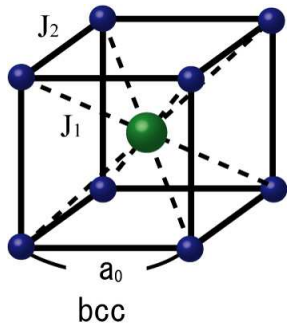}
\caption{(color online) Three-dimensional body-centered cubic (bcc)
lattices.
The filled circles denote spins connected by Heisenberg exchange interactions.
The bcc lattice is bipartite when $J_2=0$ and can be divided into two sublattices,
which are distinguished by the size of spheres. The lattice constant $a_0$ is assumed to be $1$.
\label{Fig;cubicbcc}}
\end{center}
\end{figure}

As one of the simplest models to which the general analysis given in the previous section
can be applied, we hereafter study the $J_1$-$J_2$ model on the bcc lattice
shown in Fig.~\ref{Fig;cubicbcc},
to concretely understand what kind of phases appears under high magnetic field.
Before studying a quantum case, we briefly review the ground-state
properties in the classical case in this section.

\begin{figure}[tbp]
\begin{center}
\includegraphics[scale=0.28]{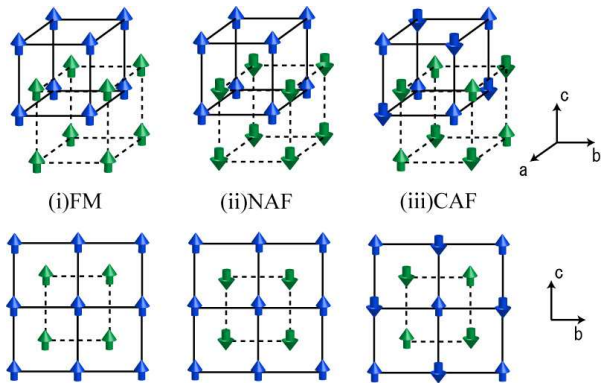}
\caption{(color online) Spin configurations for the three phases (`FM', `NAF' and `CAF')
on the bcc lattice.
(i) `FM' represents the fully polarized ferromagnetic phase.
(ii) In `NAF', the spins on each sublattice align
ferromagnetically while two spins on the different sublattices are anti-parallel.
(iii) `CAF' is composed by two antiferromagnetically-ordered sublattices,
which, as a whole, align in a collinear manner.
\label{Fig;config}}
\end{center}
\end{figure}

\begin{figure}[tb]
\begin{center}
\includegraphics[scale=0.45]{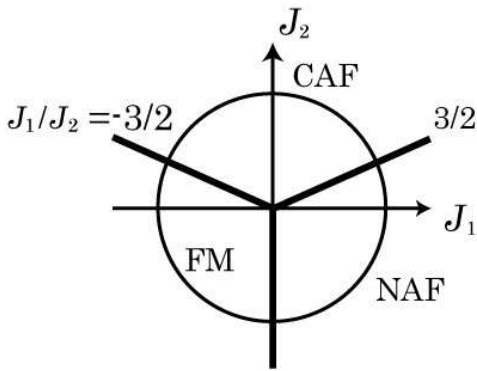}
\caption{Phase diagram of the classical $J_1$-$J_2$ model on the bcc lattice.
$J_1^2+J_2^2=1$ on the circle. The spin configurations in each phase are given
in Fig.~\ref{Fig;config}.
\label{Fig;phaseD}}
\end{center}
\end{figure}

In the absence of external field $H$, one can easily find the
classical ground state by minimizing the Fourier transform of the exchange interactions
\begin{equation}
\begin{split}
\epsilon(\bolq)&=\sum_{j} \frac{1}{2} J_{ij}
\cos [ \bolq {\cdot}(\bolr_i-\bolr_j) ]\\
&=4J_1\cos\frac{q_x}{2}\cos\frac{q_y}{2}\cos\frac{q_z}{2}\\
&+J_2(\cos q_x+\cos q_y+\cos q_z)\ ,
\end{split}
\label{epsilon}
\end{equation}
where the summation is taken over all sites connected to
site $i$ by the exchange $J_{ij}$.
At zero field, the following three types of phases appear in the ground state:
(i) a ferromagnetic phase with the wave vector $(0,0,0)$ for $J_1/J_2 < -3/2$ and $J_1<0$,
(ii) a N\'eel antiferromagnetic (NAF) phase with the wave vector $(2\pi,2\pi,2\pi)$ for $J_1/J_2 > 3/2$ and $J_1>0$,
(iii) a collinear antiferromagnetic (CAF) phase with the wave vector $(\pi,\pi,\pi)$
on each sublattice for $-3/2 < J_1/J_2 < 3/2$ and $J_2>0$, as shown in Fig.~\ref{Fig;config}.
The classical phase diagram is given in Fig.~\ref{Fig;phaseD}.
The energies of three phases are, respectively, given by
\begin{subequations}
\begin{align}
\frac{E_{\rm FM}}{NS^2}&=4J_1+3J_2,\\
\frac{E_{\rm NAF}}{NS^2}&=-4J_1+3J_2,\\
\frac{E_{\rm CAF}}{NS^2}&=-3J_2,
\label{ESF}
\end{align}
\end{subequations}
where $N$ is the number of lattice sites and $S$ denotes the length of the classical spins.

As the classical collinear antiferromagnetic ground state is not an eigenstate of the quantum model,
the argument in Sec.~II concludes that the magnetization process in the collinear antiferromagnetic
phase near
the boundary to the ferromagnetic phase at $J_1/J_2=-3/2$ must show
a nontrivial quantum behavior in the quantum model.
On the other hand, at the ferromagnetic/NAF phase boundary ($J_1=0$, $J_2<0$),
the classical NAF state is an eigenstate even in the quantum model and hence
quantum fluctuations are not important.

The classical ground state of the collinear antiferromagnetic phase 
has a non-trivial continuous degeneracy.
To see this degeneracy, let us divide the bcc lattice into two sublattices, as
shown in Fig.~\ref{Fig;cubicbcc}.
In the collinear antiferromagnetic phase,
the spins form the antiferromagnetic structure with the wave vector $(\pi,\pi,\pi)$
on each sublattice. 
The ground-state manifold of this phase has extra continuous 
degeneracy---the angle of spins between two sublattices can vary without changing the energy.
If quantum fluctuation is taken into account, the spins align in a collinear manner
because of ``order by disorder" mechanism.\cite{Shender,Henley}
When the external field is applied in the collinear antiferromagnetic phase,
the spins gradually point upward from the plane perpendicular to the external field.
There still remains this extra degeneracy in the spin components perpendicular to the field
even in the magnetization process of the classical model.
In a quantum case, the linear-spin-wave theory will predict the canted coplanar
antiferromagnetic spin state.

\begin{figure}[tb]
\begin{center}
\includegraphics[scale=0.34]{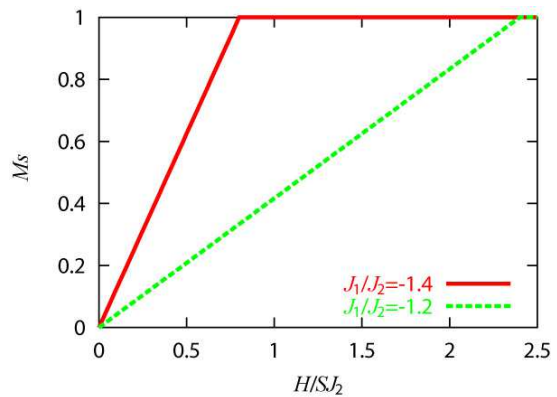}
\caption{(color online)  Magnetization curves in the classical $J_1$-$J_2$ model on the
bcc lattice with ferromagnetic $J_1<0$ and antiferromagnetic $J_2>0$.
Magnetization $M_s$
is given by $-(1/SN)\sum_\boll^{N} \VEV{S^z_\boll}$.
The saturation field $\text{H}_{c1}$ is given in Eq.\ (\ref{classical_s_field}).
\label{Fig;magMF}}
\end{center}
\end{figure}

The magnetization curve can be obtained within the mean field approximation,
by replacing the spin vector operators ${\bf S}_i/S$ with a unit vector
${\bf u}_i=(u^x_i,u^y_i,u^z_i)$.
Under the assumption of two sublattice structure, the ground-state energy of
the canted antiferromagnetic phase is given by
\begin{equation}
\begin{split}
\frac{E_{2}^\text{mean}}{(N/2)S^2}&=8J_1 u^z_A u^z_B +6 J_2[(u^z_A)^2+(u^z_B)^{2}-1]\nonumber\\
&
+\frac{\text{H}}{S}(u^z_A +u^z_B ).
\end{split}
\end{equation}
Numerically minimizing the energy with respect to $u^z_A$ and $u^z_B$,
we obtain the magnetization process shown in  Fig.~\ref{Fig;magMF}.
With increasing external magnetic field, the magnetization increases uniformly
and saturates at the classical saturation field $\text{H}_{c1}$,
\begin{equation}
{\cal H}_{c1}=2S[\epsilon({\bf 0})-\epsilon_{\text{min}}],
\label{classical_s_field}
\end{equation}
where $\epsilon_{\text{min}}$ denotes the minimum value of dispersion $\epsilon(\bolq)$.


\section{various possible phases from the single-magnon BEC}
\label{Sec:singleBEC}
Since the classical ground state of the bcc $J_1$-$J_2$ model
has a large degree of degeneracy in the parameter range $-3/2 < J_1/J_2 < 3/2$,
it is expected that quantum fluctuations would single out a certain phase
due to the mechanism of ``order by disorder".
Slightly below the saturation field, the method of a dilute Bose gas\cite{Batyev} enables us
to predict non-biased results of the nature of single-magnon BEC in a fully quantum manner.
In this section, we discuss the various possible phases emerging from
the single-magnon condensation (BEC) near the saturation
in the $S=1/2$ bcc $J_1$-$J_2$ model in the range
$-3/2 < J_1/J_2 < 3/2$.

In the hardcore boson language,
the spin operators at site $l$ are written as\cite{Matsubara-Matsuda}
\begin{equation}
S^z_l=-1/2+a^\dagger_la_l,\ \ \ S_l^+ =a_l^\dagger,\ \ \ S_l^- =a_l
\label{Ch1:hardcoreSpin}
\end{equation}
near saturation, where $a_l$ denotes the annihilation operator of a boson (magnon).
The vacuum $\ket{\Omega}$ of bosons, i.e.\ $a_l\ket{\Omega}=0$, corresponds
to the saturated ferromagnetic state $\ket{\Omega}=\otimes_j |\downarrow\rangle_j$.
In terms of these boson operators, the Hamiltonian reads
\begin{align}
{\cal H} &= \sum_{q}[\omega (\bolq) - \mu]a^\dagger_{\bolq}a_{\bolq}
+\frac{1}{2N}  \sum_{\bolq,\bolk,\bolk^\prime}  V_{\bolq}
a_{\bolk+\bolq}^\dagger a_{\bolk^\prime-\bolq}^\dagger
a_{\bolk}a_{\bolk^\prime},
\label{Hboson}
\end{align}
where
\begin{align}
\omega(\bolq) &=\epsilon(\bolq)-\epsilon_{\text{min}}\ ,
\nonumber\\
\mu&={\rm H}_{c1}-{\rm H},\nonumber\\
V_{\bolq} &=2[\epsilon(\bolq)+U].
\end{align}
The interaction $U$ is the on-site hard-core potential, which
is set to infinity $U{\rightarrow}\infty$ afterward.
For $-3/2<J_1/J_2<3/2$, the magnon dispersion $\epsilon(\bolq)$ takes its minimum
$\epsilon_{\text{min}}=-3J_2$ at two wave vectors $\pm\bolQ$, where $\bolQ=(\pi,\pi,\pi)$.

When external field is lower than the saturation field ($\text{H} < \text{H}_{\text{c}1}$
or equivalently $\mu>0$),
BEC of magnons can occur in two momenta:
\begin{align}
\VEV{a_{\bolQ}}&=\sqrt{N\rho_{\bolQ}}\exp(i\theta_{\bolQ}),\\
\VEV{a_{-\bolQ}}&=\sqrt{N\rho_{-\bolQ}}\exp(i\theta_{-\bolQ}).
\end{align}
The induced spin-ordered phase is characterized by the wave vectors $\bolQ$ and/or $-\bolQ$.

The effective energy per site $E/N$ of the dilute Bose gas
is determined by the interaction between the bosons condensed at ${\bf q}=\pm {\bf Q}$.
In the dilute limit, the energy density $E/N$ is expanded with the density
$\rho_{\pm {\bf Q}}$
up to quadratic terms in the form
\begin{align}
%
\frac{E}{N} =&\frac{\Gamma_1}{2}
\left(\rho_{\bolQ}^2+\rho_{-\bolQ}^2\right)
+[ \Gamma_2 +\Gamma_3\cos 2(\theta_\bolQ-\theta_{-\bolQ}) ]
\rho_{\bolQ} \rho_{-\bolQ} \nonumber\\
&- \mu(\rho_{\bolQ}+\rho_{-\bolQ}).
\label{Ch2:EffPotential}
\end{align}
Here we introduced the renormalized interaction $\Gamma_1$ between
bosons with the same mode, $\Gamma_2$ between bosons with the different modes,
and $\Gamma_3$ obtained from an umklapp scattering.

\begin{figure}[bt]
\begin{center}
\includegraphics[scale=0.34]{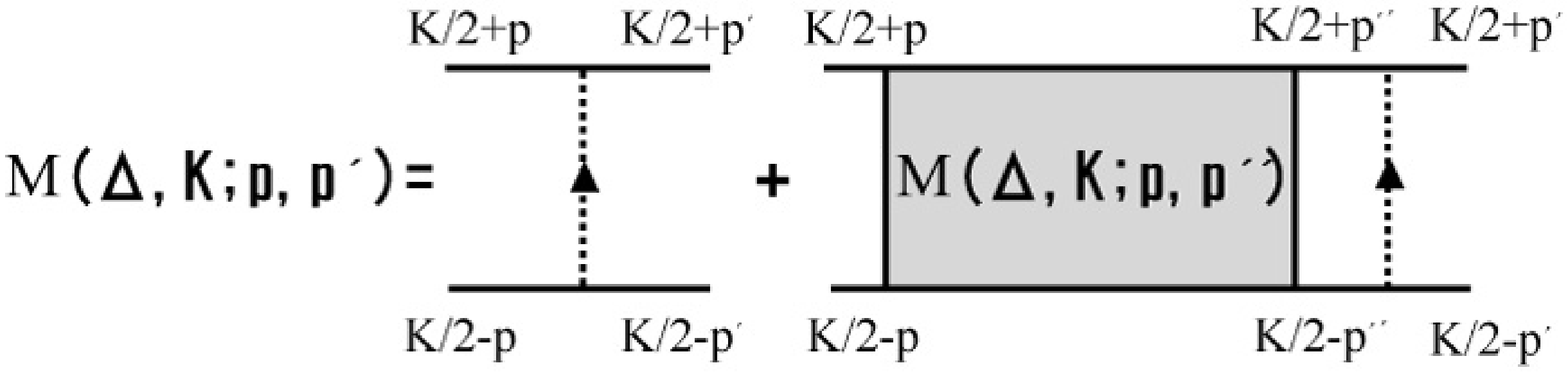}
\caption{\label{fig:scattering}
Scattering amplitude $M$ given by the ladder diagram.
\label{Fig:scatteringM}}
\end{center}
\end{figure}

More explicitly, the renormalized interactions $\Gamma_\mu$ ($\mu=1,2,3$) are
exactly obtained
by the scattering amplitude $M$ of two magnons.
In the case of two magnons on the saturated ferromagnetic state, $M$ is exactly given by the ladder
diagram (shown in Fig.~\ref{Fig:scatteringM}) in the form
\begin{equation}
\begin{split}
&M(\Delta,\bolK;\bolp,\bolp^\prime)=V_{\bolp^\prime-\bolp}+V_{-\bolp^\prime-\bolp}\\
&-\frac{1}{2}\int \frac{d^3 p^{\prime\prime}}{(2\pi)^3}
\frac{M(\Delta,\bolK;\bolp,\bolp^{\prime\prime})
(V_{\bolp^\prime-\bolp^{\prime\prime}}+V_{-\bolp^\prime-\bolp^{\prime\prime}})}
{\omega(\bolK/2+\bolp^{\prime\prime})+\omega(\bolK/2-\bolp^{\prime\prime})+\Delta-i0^+},
\label{laddereq}
\end{split}
\end{equation}
where $\Delta$ is the total energy and $\bolK$ is the center-of-mass momentum of two magnons.
Using the scattering amplitude, $\Gamma_\mu$ are given as
\begin{equation}
\begin{split}
\Gamma_1 &=M(0,2\bolQ;0,0)/2\ ,\\
\Gamma_2 &=M(0,0;\bolQ,\bolQ)\ ,\\
\Gamma_3&=M(0,2\bolQ;0,\bolq_1)/2\ ,
\end{split}
\label{Gam3}
\end{equation}
where $\bolq_1=(0,0,2\pi)$.
We note that our expression is different from the previous ones\cite{Batyev,Nikuni-Shiba-2,J1J2magBEC}
but can be easily recast into them.
The scattering amplitude of this form is appropriate
to study the bound state in terms of the Bethe-Salpeter equation\cite{BetheSalpeterReview};
a divergence of the $M$ implies a stable bound state respecting the permutation symmetry of bosons.
We will review how to solve Eq.~(\ref{laddereq}) and how to obtain
the properties of bound state in Appendix.~\ref{Ap:bm}.

Before calculating these values concretely,
let us discuss the possible emergent phases depending on the values of $\Gamma_{i}$.
The densities $\rho_{\pm {\bf Q}}$ of condensed bosons
 in the ground state are obtained by minimizing the energy $E/N$.
The phases $\theta_{\pm \bolQ}$ are pinned as
$\theta_\bolQ-\theta_{-\bolQ}=(m+\frac12)\pi$ and $m\pi$ with an integer $m$,
respectively,
in the case that $\Gamma_3$ is positive and negative.

\subsection{Canted coplanar antiferromagnetic phase}
When the renormalized interactions satisfy the condition $\Gamma_1>\Gamma_2-|\Gamma_3|>-\Gamma_1$,
the two modes condense simultaneously with the same density
given by
$\rho_{\bolQ} =\rho_{-\bolQ}=\rho=\mu/(\Gamma_1+\Gamma_2-|\Gamma_3|)$
in the ground state and hence
\begin{equation}
\begin{split}
\VEV{a_l}& =\sqrt{\rho}
\left[
\be^{i({\bf Q}{\cdot}{\bf R}_l+\theta_\bolQ)}
+\be^{i(-{\bf Q}{\cdot}{\bf R}_l+\theta_{-\bolQ})}
\right].
\end{split}
\end{equation}
In this case, the spin expectation values are given as
\begin{equation}
\begin{split}
\VEV{S_l^z}
 &=-\frac{1}{2}+4\rho\cos^2 \left({\bf Q}{\cdot} {\bf R}_l
+\frac{\theta_{\bolQ}-\theta_{-\bolQ}}{2}\right),\\
\VEV{S_{l}^{\pm}}
 &=2\sqrt{\rho}\cos \left({\bf Q}{\cdot} {\bf R}_l
+\frac{\theta_\bolQ-\theta_{-\bolQ}}{2}\right)
\be^{\mp i\theta},
\label{Ch2:fan}
\end{split}
\end{equation}
where we set
$\theta\equiv (\theta_\bolQ+\theta_{-\bolQ})/2$.
In the case $\Gamma_3>0$, the phases $\theta_{\bolQ}$ and $\theta_{-\bolQ}$
are pinned as $\theta_{\bolQ}-\theta_{-\bolQ}=(m+\frac12)\pi$.
Setting $m=0$, we thus obtain
\begin{equation}
\begin{split}
\VEV{S_l^z}
 &=-\frac{1}{2}+2\rho,\\
\VEV{S_{l}^{\pm}}
 &=2\sqrt{\rho}\cos \left({\bf Q}{\cdot} {\bf R}_l
+\frac{\pi}{4}\right)\be^{\mp i\theta},
\end{split}
\end{equation}
where the factor
$2\cos \left({\bf Q}{\cdot} {\bf R}_l
+\frac{\pi}{4}\right)=1$ or $-1$ on the bcc lattice.
The spin $xy$ components have an antiferromagnetic
order which has the same sublattice structure as the classical collinear antiferromagnetic state
shown in Fig.~\ref{Fig;config}(iii).

\subsection{Spin-supersolid phase}
When $\Gamma_1>\Gamma_2-|\Gamma_3|>-\Gamma_1$ and $\Gamma_3<0$,
the two modes condense simultaneously and the spin expectation values satisfy Eq.~(\ref{Ch2:fan}).
In contrast to the canted antiferromagnetic phase, the negative $\Gamma_3$ leads to
$\theta_{\bolQ}-\theta_{-\bolQ}=m\pi$.
In the case $m=0$, the emergent phase has the following spin expectation values
\begin{equation}
\begin{split}
\VEV{S_l^z}
&=-\frac{1}{2},\ \ \
\VEV{S_{l}^{\pm}}=0
\label{SS1}
\end{split}
\end{equation}
on $A$ sublattice, defined in Fig.~\ref{Fig;cubicbcc}, and
\begin{equation}
\begin{split}
\VEV{S_l^z}
 &=-\frac{1}{2}+4\rho,\\
\VEV{S_{l}^{\pm}}
 &=2\sqrt{\rho}\cos ({\bf Q}{\cdot} {\bf R}_{l})e^{\mp i \theta}
\label{SS}
\end{split}
\end{equation}
on $B$ sublattice. Here $A$ sublattice includes the site at ${\bf R}=0$.
For an odd integer $m$, $A$ and $B$ sublattices interchange.
In this phase, the density of boson ($\VEV{S_l^z}$) oscillates, breaking
the translational symmetry spontaneously. 
Together with the spin density wave,
the spin $xy$ component on B sublattice also has an antiferromagnetic N\'eel order.
Hence, this condensed phase can be considered as a spin-supersolid phase.

\subsection{Chirality-breaking spiral phase}
When $\Gamma_2-|\Gamma_3| > \Gamma_1 >0$,
a density difference between two modes appears and
the doubly-degenerate ground states are given by
$\rho_{\bolQ} =\rho^\prime=\mu/\Gamma_1$ and
$\rho_{-\bolQ}=0$, and   vice versa.
When the magnons with the wave vector $\bolQ$ are condensed as
\begin{equation}
\VEV{a_l}=\sqrt{\rho^\prime}\exp[i({\bf Q}{\cdot}{\bf R}_l+\theta_{\bolQ})],
\end{equation}
the spin expectation values are explicitly described as
\begin{equation}
\begin{split}
& \VEV{S_l^z}=-\frac{1}{2}+\rho^\prime\ ,\\
& \VEV{S_l^x} = \sqrt{\rho^\prime}\cos ({\bf Q}{\cdot}{\bf R}_l + \theta_{\bolQ})\ ,\\
& \VEV{S_l^y} = -\sqrt{\rho^\prime}\sin ({\bf Q}{\cdot}{\bf R}_l + \theta_{\bolQ}).\\
\end{split}
\end{equation}
For the condensation with the wave vectors $\pm \bolQ$,
spin $xy$ components form a spiral structure as
$\VEV{S_l^y}/\VEV{S_l^x}=\mp \tan ({\bf Q}{\cdot}{\bf R}_l + \theta_{\pm \bolQ})$,
in the direction of $(1,\pm 1,\pm 1)$, where the pitch angle of spiral is $\pm\pi/2$.
This phase spontaneously breaks the chiral symmetry,
so that a multiferroic behavior is expected to be
accompanied.\cite{multiferro_1,multiferro_2,multiferro_3}

\subsection{Phase separation}


When one of the effective interactions is negative, $\Gamma_1<0$ or $\Gamma_1+\Gamma_2-|\Gamma_3|<0$,
the low-energy bosons at $\mathbf{q}=\pm \mathbf{Q}$
attract each other, which makes magnon condensed states in the low-density limit unstable.
It is natural to expect
a first order transition, or equivalently, phase separation between the fully polarized state
and a low-magnetization state.

When $\Gamma_1+\Gamma_2-|\Gamma_3|$ is negative,
the form of energy density $E/N$ [see Eq.~(\ref{Ch2:EffPotential})]
indicates that BEC of two modes with
an equal high density always has a lower energy
than other low-density states.
In this case, if  $\Gamma_3>0$ ($\Gamma_3 <0$), we expect the occurrence of a phase separation
between the fully polarized state and
a canted coplanar antiferromagnetic state (spin-supersolid state) with a low magnetization,
which is accompanied with a magnetization jump
below the saturation field in the magnetization curve.

When $\Gamma_1$ is negative and satisfies $\Gamma_1<\Gamma_2-|\Gamma_3|$,
the energy density $E/N$ [see Eq.~(\ref{Ch2:EffPotential})]
indicates that BEC of a single mode with a certain high density always has a lower energy
than other low-density states.
In this case, we hence expect the appearance of a phase separation between the fully polarized state
and a spiral state with a low magnetization, accompanied by a magnetization jump.


So far, our argument based on Eq.~(\ref{Ch2:EffPotential}) is restricted to
the magnon BEC that occurs in the single-particle channel.
However, the strong attraction can also induce formation of magnon bound states,
which lead to the BEC of bound multi magnons.
This possibility is examined in Sec.~VI.

\subsection{Degenerate case}
When the renormalized interactions satisfy a special condition
$\Gamma_1 = \Gamma_2-|\Gamma_3| >0$,
the ground state has an infinite degeneracy, given by
$\rho=\mu/\Gamma_1$,
$\rho_{\bolQ} =\rho\cos^2\theta_d$, and
$\rho_{-\bolQ}=\rho\sin^2\theta_d$ for arbitrary $\theta_d$.

\subsubsection{Case of $\Gamma_3>0$}
In the case $\Gamma_3>0$, the phases are pinned as
$\theta_{\bolQ}-\theta_{-\bolQ}=\pi/4$.
The spin expectation values in the degenerate ground states are given as
\begin{equation}
\begin{split}
&\VEV{S_l^z}=-\frac{1}{2}+\rho,\\
& \VEV{S_l^x} = \sqrt{\rho}\cos ({\bf Q}{\cdot}{\bf R}_l +\theta_{\bolQ}-\theta_d),\\
& \VEV{S_l^y} = -\sqrt{\rho}\sin ({\bf Q}{\cdot}{\bf R}_l +\theta_{\bolQ}-\theta_d),
\end{split}
\label{Inf:Deg1A}
\end{equation}
for $A$ sublattice and
\begin{equation}
\begin{split}
&\VEV{S_l^z}=-\frac{1}{2}+\rho,\\
& \VEV{S_{l}^x} = \sqrt{\rho}\cos ({\bf Q}{\cdot}{\bf R}_{l} +\theta_{\bolQ}+\theta_d),\\
& \VEV{S_{l}^y} = -\sqrt{\rho}\sin ({\bf Q}{\cdot}{\bf R}_{l} +\theta_{\bolQ}+\theta_d),
\end{split}
\label{Inf:Deg1B}
\end{equation}
for $B$ sublattice.
Hence, $\theta_d$ controls the relative angle in the $x$-$y$ plane between spins
on A and B sublattices.
In the classical limit of $S\rightarrow \infty$,
this infinite degeneracy is realized as discussed in Appendix.~\ref{Ap:classical}.

\subsubsection{Case of $\Gamma_3<0$}
In the case $\Gamma_3<0$, we have $\theta_{\bolQ}-\theta_{-\bolQ}=0$ and
obtain
\begin{equation}
\begin{split}
&\VEV{S_l^z}=-\frac{1}{2}+\rho[1+\sin 2\theta_d\cos(2{\bf Q}{\cdot}{\bf R}_l)]\ ,\\
& \VEV{S_l^x} = \sqrt{2\rho}
\cos\left(\theta_d-\frac{\pi}{4}\right)\cos ({\bf Q}{\cdot}{\bf R}_l +\theta_{\bolQ}),\\
& \VEV{S_l^y} = \sqrt{2\rho}\sin\left(\theta_d-\frac{\pi}{4}\right)
\sin ({\bf Q}{\cdot}{\bf R}_l +\theta_{\bolQ}).
\end{split}
\end{equation}
Now, the spin-supersolid phase appears and the modulation of the density ($\VEV{S_l^z}$) depends on $\theta_d$.

\section{phase diagram of the single-magnon BEC}
\label{sec:phaseD}
In this section, we determine the magnetic structures
in the high-magnetic-field regime
of the $S=1/2$ $J_1$-$J_2$ model on the bcc lattice, in the parameter range
$-3/2 \leq J_1/J_2 \leq 3/2$ with $J_2>0$.
We only consider spin structures induced by
single magnon BECs in this section.
A possible bound magnon BEC is also considered in Sec.~\ref{sec:BMBEC}.

\begin{figure}[tb]
\begin{center}
\includegraphics[scale=0.35]{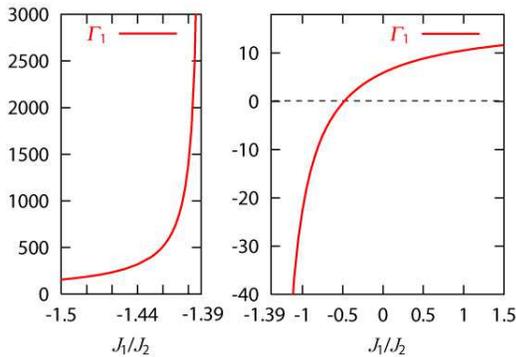}
\caption{(color online) Interaction $\Gamma_1$ plotted as a function of $J_1/J_2$ for $J_2>0$.
$\Gamma_1$ diverges at $J_1/J_2=-1.389$.
\label{Gamma11}
}
\end{center}
\end{figure}

\begin{figure}[tb]
\begin{center}
\includegraphics[scale=0.35]{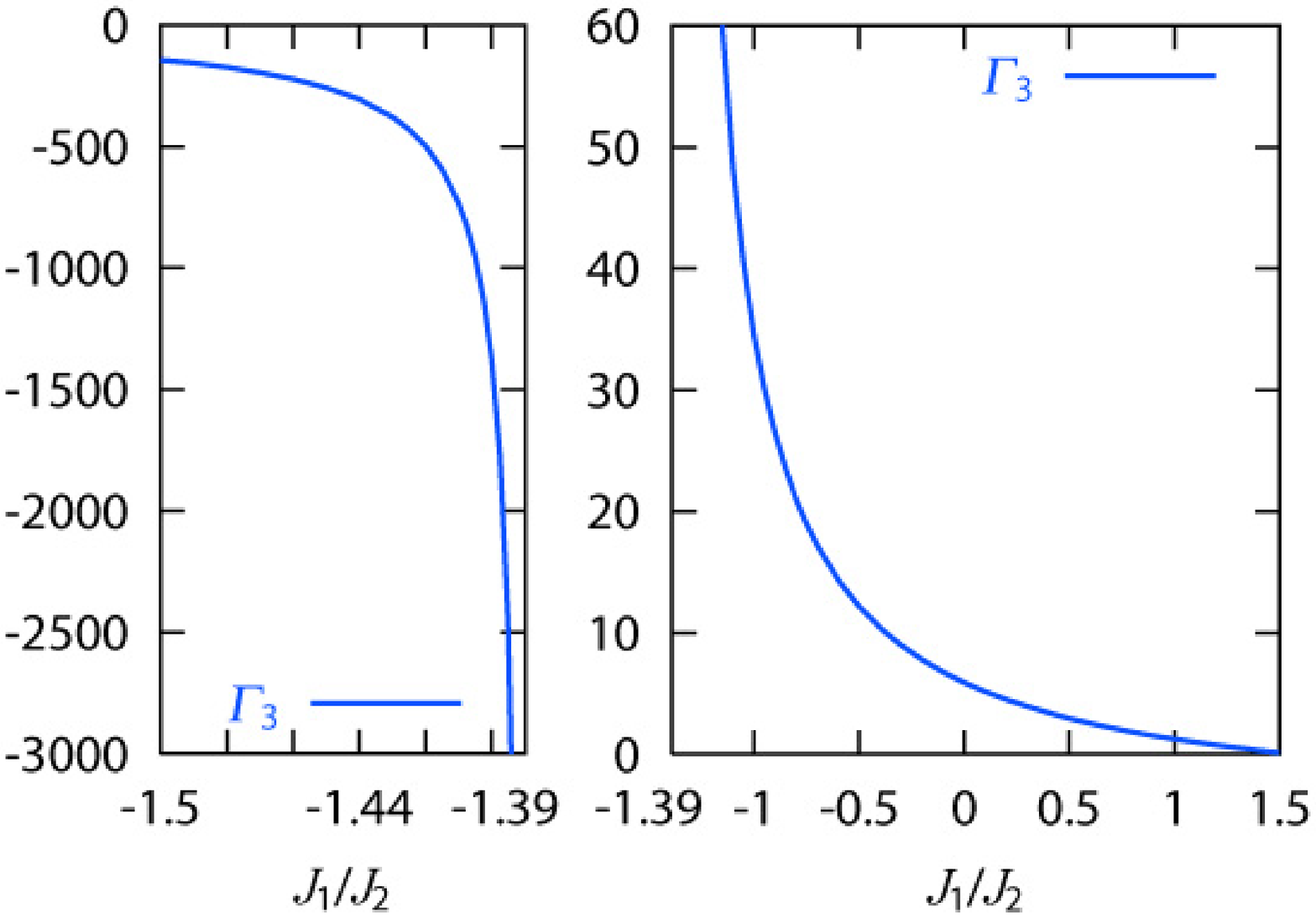}
\caption{(color online) Interaction $\Gamma_3$ plotted as a function of $J_1/J_2$ for $J_2>0$.
$\Gamma_3$ diverges at $J_1/J_2=-1.389$.
\label{fig:Gamma3}
}
\end{center}
\end{figure}

\begin{figure}[tb]
\begin{center}
\includegraphics[scale=0.35]{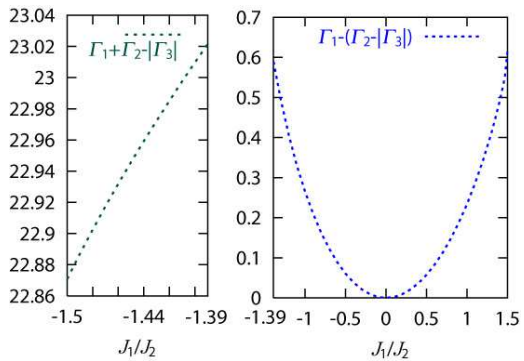}
\caption{(color online) Effective interactions $\Gamma_1+\Gamma_2-|\Gamma_3|$ and
$\Gamma_1-(\Gamma_2-|\Gamma_3|)$ plotted as a function of $J_1/J_2$ for $J_2>0$.
\label{Gamma123}
}
\end{center}
\end{figure}

Using the method described in Appendix.~\ref{Ap:bm},
we numerically calculate the interactions $\Gamma_i$ ($\mu=1,2,3$).
The results are shown
in Figs.~\ref{Gamma11}, \ref{fig:Gamma3}, and \ref{Gamma123}. From
the energy comparison described in the preceding section,
we find that the following three phases appear
slightly below the saturation field:

(i) Canted coplanar antiferromagnetic phase: in the parameter range $-0.48 \leq J_1/J_2 \leq 3/2$,
a canted coplanar antiferromagnetic phase appears since the interactions satisfy
$\Gamma_1>\Gamma_2-|\Gamma_3|>-\Gamma_1$ and $\Gamma_3>0$.

(ii) Phase separation: in the range $-1.389 \leq J_1/J_2 \leq -0.48$,
we obtain $\Gamma_1<0$ and hence
a phase separation is expected below the saturation field.
The resulting phase below the magnetization jump is plausibly the canted coplanar
antiferromagnetic phase
since the interactions satisfy $\Gamma_1-(\Gamma_2-|\Gamma_3|)>0$ in this parameter range.
We do not exclude the possibility of other nontrivial phases
such as the spin nematic phase appearing through a 1st-order transition.
At $J_1/J_2=-1.389$, both $\Gamma_{1}$ and $\Gamma_{3}$
diverge, which implies the appearance of bound states.

(iii) Spin supersolid phase:
for $-3/2\leq J_1/J_2 \leq -1.389$,
we have $\Gamma_1>\Gamma_2-|\Gamma_3|>-\Gamma_1$
and $\Gamma_3<0$, which suggests the appearance of a spin supersolid phase.
However, in this parameter range, we find that
a two-magnon bound state already has negative energy at the saturation field
given by the one-magnon flip.
Condensation of stable two-magnon bound states
leads to a spin nematic phase.
In the next section,
we take account of the possibility of this bound-two-magnon BEC.
It should be also noted that the negativeness of $\Gamma_3$ also
requires caution.
A stable two-magnon bound state results in the appearance of a pole
in the two-magnon propagator, which changes the sign of interaction $\Gamma_i$
when the energy level crosses the pole. Hence the value of $\Gamma_3$
is strongly influenced by the presence of the pole in this parameter range.

\section{Bound-magnon BEC}
\label{sec:BMBEC}
In sufficiently high fields, the magnon dispersion has a gap on the fully polarized state.
When two-magnon bound states have lower energy than single magnons,
the energy gap of two-magnon bound states closes earlier than
that of single magnons, with decreasing magnetic field.
Below the saturation field, bound magnon pairs condense,
forming a spin nematic state.\cite{SquareNem}
The bound-magnon-condensed phases
have different properties from the single-magnon BEC.
The striking one is the absence of the transverse local magnetization,
and instead the existence of the long range order in the quadratic channels
$\VEV{S_i^+S_j^+}\neq 0$ for a certain bond $(i,j)$,
which corresponds to the spin nematic order.\cite{Chubukov,SquareNem}
The existence of stable two-magnon-bound states gives rise to divergence
of the scattering amplitude of two magnons.
In this section, we mainly focus on the analysis of obtained results.
The detailed method for the calculations is given in Appendix.~\ref{Ap:bm}.

\begin{figure}[tb]
\begin{center}
\includegraphics[scale=0.5]{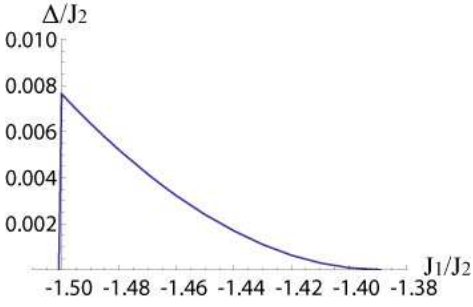}
\caption{(color online) Binding energy $\Delta_B$ of magnon pairs
with the center-of-mass momentum $\bolK_0=(2\pi,2\pi,2\pi)$.
In the case of positive $\Delta_B$, the bound-magnon BEC is expected.
\label{Ch3:bccDelta}
}
\end{center}
\end{figure}

To see the possibility of a bound magnon BEC,
we study the binding energy 
of two-magnon bound states.
We find that the minimum point of the energy dispersion
$-\Delta_B(\bolK)$
of bound magnon pairs
always exists at $\bolK=\bolK_0=(2\pi,2\pi,2\pi)$. From
Eq.~(\ref{eq:b_energy}),
the binding energy $\Delta_B=\Delta_B(\bolK_0)$ is given by solving
\begin{equation}
J_1\int \frac{d^3 p}{\pi^3}\frac{\cos ^2\frac{p_x}{2}\cos^2\frac{p_y}{2}\cos^2\frac{p_z}{2}}
{\omega(\bolK_0/2+\bolp)+\omega(\bolK_0/2-\bolp)-\Delta_B}=1,
\end{equation}
where
\begin{align}
&\omega(\bolK_0/2+\bolp)+\omega(\bolK_0/2-\bolp)\nonumber\\
&= \left\{ \begin{array}{l}
           8J_1+2J_2(3+\cos p_x +\cos p_y + \cos p_z) \\
           \hspace{3.5cm}\mbox{(for $J_1/J_2\leq -3/2$)}, \\
           6J_2-2J_2(\cos p_x +\cos p_y + \cos p_z) \\
           \hspace{3.5cm}\mbox{(for $J_1/J_2\geq -3/2$)}.
         \end{array}
         \right.
\end{align}
The total bound-magnon
energy is given by $E_B=-2\mu-\Delta_B$.



\begin{figure}[tb]
\begin{center}
\includegraphics[scale=0.67]{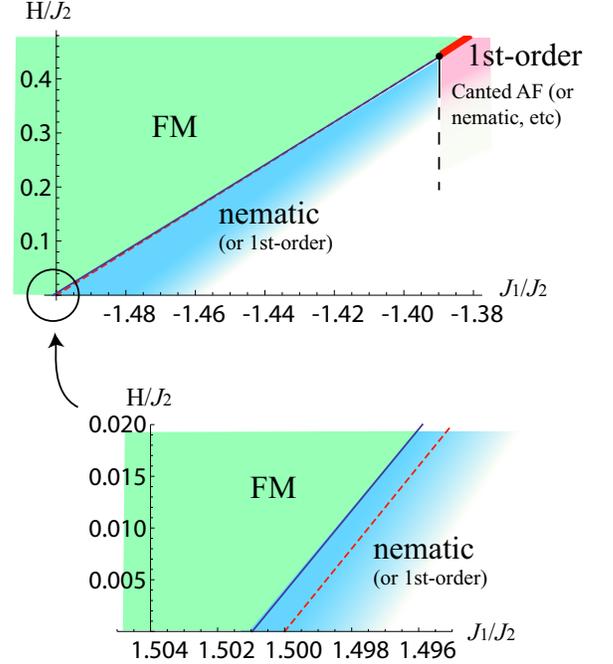}
\caption{(color online) Schematic phase diagram near
the critical external field ${\rm H}_c$.
For ${\rm H}>{\rm H}_c$, the spins form the saturated ferromagnetic state.
The straight line (blue) denotes the critical field where the gap of
bound magnon pairs closes.
The short dashed line and the bold one (red)
denotes the field where the single-magnon gap closes.
In particular, at the bold line, the phase separation occurs below the saturation field.
For $-1.50097\leq J_1/J_2\leq -1.389$,
the gap of the bound magnon closes earlier than that of the single magnon,
and the spin nematic phase is expected to appear.
`1st-order' corresponds to the phase separation (magnetization jump).
The vertical dashed line is a schematic phase boundary
between the bound magnon BEC and the single magnon BEC.
\label{Ch3:CriticalHBcc}
}
\end{center}
\end{figure}
The obtained binding energy $\Delta_B$ and the critical external field ${\rm H}_c$
are, respectively, shown in Figs.~\ref{Ch3:bccDelta} and \ref{Ch3:CriticalHBcc}.
In the range $-3/2\leq J_1/J_2\leq -1.389$, the gap of the bound-magnon pairs above
the saturation closes earlier than
that of the single magnons with lowering the magnetic field. Hence
a spin-nematic phase appears below the saturation field,
which has a lower energy than the spin supersolid phase discussed in Sec.~\ref{sec:phaseD}.
Even in the range $-1.50097\leq J_1/J_2\leq -3/2$, the ground state
is not the classically expected ferromagnetic phase in the absence of external field,
since the reference ferromagnetic state is destabilized by the fluctuation of the bound-magnon pairs.
The FM/nematic phase transition in zero field
may be 1st-order due to a substantial difference in magnetization, similar to
the case of the square lattice.\cite{SquareNem}



Here, we analyze the properties of the bound magnon condensed phase.
The wave function of the bound-magnon pair is given by
\begin{equation}
\sum_\bolp \chi_{\bolK_0}(\bolp)S^+_{\bolp}S^+_{\bolK_0-\bolp}
|\text{FM}\rangle
\end{equation}
with
\begin{equation}
\chi_{\bolK_0} (\bolp)\propto \frac{\sin\frac{p_x}{2}\sin\frac{p_y}{2}\sin\frac{p_z}{2}}
{\omega(\bolp)+\omega(\bolK_0-\bolp)-\Delta_B},
\end{equation}
where $|\text{FM}\rangle=\otimes_j|\downarrow\rangle_j$ and we have omitted the normalization constant.
Note that we have translated the momentum $\bolp$ of $\chi_{\bolK_0} (\bolp)$
by $\bolK_0/2$ in comparison with the definition in Appendix B2
to see the symmetry of the bound state clearly.
The symmetry of this wave function $\chi(\bolp)$ corresponds to the space symmetry of
spin nematic order.\cite{SquareNem}
Hence, the nematic order parameter, which is dominant on the nearest neighbor bonds,
has the $f_{xyz}$-wave symmetry and oscillates with the wave vector $\bolK_0$ in the form
\begin{align}
&\VEV{S^+_iS^+_{i\pm( {\bf e}_x+{\bf e}_y +{\bf e}_z)/2}}
=-\VEV{S^+_iS^+_{i\pm( {\bf e}_x+ {\bf e}_y - {\bf e}_z)/2}}\nonumber\\
&=\VEV{S^+_iS^+_{i\pm( {\bf e}_x- {\bf e}_y - {\bf e}_z)/2}}
=-\VEV{ S^+_iS^+_{i\pm( {\bf e}_x- {\bf e}_y+ {\bf e}_z)/2} }\nonumber\\
&=\pm \sqrt{\rho_2}\exp(i2\theta),
\end{align}
where $\rho_2$ denotes the density of condensed two-magnon bound states
and $\theta$ denotes the rotation angle of directors about $z$-axis.


Finally, let us discuss the possibility of magnetization jump
below the saturation field,
which we do not exclude from the present analysis.
Even if the magnetization jumps,
we still expect the appearance of the spin nematic phase below the magnetization jump.
It should be noted that
the spin nematic phase may seriously compete with
a single magnon BEC phase below the jump
since the binding energy is extremely small.
To investigate the full magnetization curve,
we also performed an exact diagonalization analysis of finite size systems up to $N=36$ spins.
We found a signature of magnetization jump below the saturation
and the appearance of spin nematic phase below the jump.
This tendency toward the spin nematic ordering continues down to zero
magnetization, but it is difficult to conclude the nature in the thermodynamic limit because
of the large size dependence in contrast to the small binding energy.
This issue is beyond the scope of the present paper.

\section{conclusion}
To summarize, we have discussed effects of quantum fluctuations
in quantum frustrated ferromagnets under high magnetic field.
In general, in the vicinity of the classical boundary between ferromagnetic and antiferromagnetic phases,
a non-classical phase or phase separation must appear, at least in the high-magnetic-field
regime near the saturation field,
if the classical antiferromagnetic state is not an eigenstate of the quantum Hamiltonian.
We stress that the phase separation can occur even in the isotropic Heisenberg model and
it is purely a quantum phenomenon.

As a concrete model satisfying the above condition, we study the $S=1/2$ $J_1$-$J_2$ model
on the bcc lattice, using a dilute-Bose-gas approach.
The result is summarized in Table~\ref{table:phasetransition}.
We found that for $-1.389 \leq J_1/J_2 \leq -0.48$,
the low-density single magnon condensed phase is not stable near the saturation field
and the phase separation occurs between a low-magnetization phase and the fully
polarized ferromagnetic phase, accompanied with a magnetization jump.
We also found that for $-1.50097\leq J_1/J_2\leq -1.389$ the bound magnon pairs
are stabilized, leading to a spin nematic phase.
When the density of magnons is higher than a certain amount, i.e., the magnetization is
low,
the competition between the spin nematic phase and the single magnon BEC phase might be
crucial, since the binding energy is extremely small.
It is beyond the scope of our paper to determine which phase actually appears near zero field.

Finally, let us comment on the possibility of the Efimov effect.
\cite{Efimov}
Recently, Nishida, Kato and Batista discussed
that the Efimov effect can be realized in quantum magnets
when the $s$-wave scattering length of magnons diverges.\cite{Nishida}
As shown in Fig.~\ref{Gamma11}, the $s$-wave scattering length $\Gamma_1$
diverges at $J_1/J_2=-1.389$.
Naively, in the $J_1$-$J_2$ model on the bcc lattice the Efimov effect
is also expected in the fully polarized phase.
However, in our case,
the stable bound state has a $f$-wave symmetry, instead of the $s$-wave symmetry,
which implies the scattering is not simple.
Hence, the application of the Efimov's argument to our model
may not be straightforward.
This point is also a remaining issue.


\begin{table}
\caption{Phases appearing from single-magnon and bound-two-magnon instabilities
at the saturation field in the $S=1/2$ bcc $J_1$-$J_2$ model with $J_2>0$.
For $-1.50097\leq J_1/J_2 \leq -1.389$, the two-magnon bound state is stabilized,
inducing a spin nematic phase.
}
\label{table:phasetransition}
\begin{center}
\begin{ruledtabular}
\begin{tabular}{c|cc}
\raisebox{0.5ex}[0pt]{$J_1/J_2$} &
\raisebox{0.5ex}[0pt]{single magnon} &
\raisebox{0.5ex}[0pt]{bound magnon} \\
\hline
-1.50097 $\sim$ -1.389 & -- & spin nematic \\
-1.389 $\sim$ -0.48 & phase separation/canted AF & -- \\
-0.48 $\sim$ 1.5 & canted AF & --\\
\end{tabular}
\end{ruledtabular}
\end{center}
\end{table}

\begin{acknowledgements}
The authors thank Keisuke Totsuka and Yusuke Nishida for helpful discussions.
This work was supported by
KAKENHI Grants No.~22014016 and No. ~23540397.

\end{acknowledgements}

\appendix
\section{Large S expansion}
\label{Ap:classical}
In this appendix, we study an emergent phase
near the saturation field for the large $S$ limit (classical limit).
To explicitly control $S$, we introduce the spinwave Hamiltonian using the Dyson-Maleev transformation\cite{Dyson,Oguchi}:
\begin{equation}
\begin{split}
S_\boll^+&=\sqrt{2S}a_\boll^\dagger\left(1-\frac{a_\boll^\dagger a_\boll}{2S}\right),\\
S_\boll^-&=\sqrt{2S}a_\boll,\\
S^z_\boll&=-S+a^\dagger_\boll a_\boll .
\end{split}
\label{eq;ADM}
\end{equation}
The Hamiltonian is given by
\begin{align}
&{\cal H}=\sum_\bolk [\omega_{s} (\bolq) - \mu_s] a^\dagger_\bolk a_\bolk\nonumber\\
&\ +\frac{1}{2N}\sum_{\bolk_1,\bolk_2,\bolq}
V_s(\bolq;\bolk_1,\bolk_2)a^\dagger_{\bolk_1+\bolq}a^\dagger_{\bolk_2-\bolq}
a_{\bolk_1} a_{\bolk_2},
\end{align}
where
\begin{align}
&\omega_s(\bolq) =2S[\epsilon(\bolq)-\epsilon_{\text{min}}],\\
&V_s(\bolq;\bolk_1,\bolk_2) =2\epsilon(\bolq)-\epsilon(\bolk_1)-\epsilon(\bolk_2),
\end{align}
$\mu_s={\rm H}_{c1}-{\rm H}$, and ${\rm H}_{c1}$ is given in Eq.\ (\ref{classical_s_field}).
In first order in $S^{-1}$, the renormalized interactions $\Gamma_\mu$
in Eq.~(\ref{Ch2:EffPotential}) are given by
\begin{subequations}
\begin{align}
\Gamma_1 &=V(0;\bolQ,\bolQ)=8J_1+12J_2\ ,\\
\Gamma_2 &=V(0;-\bolQ,\bolQ)+V(2\bolQ;-\bolQ,\bolQ)=24J_2\ ,\\
\Gamma_3 &=V(2\bolQ;\bolQ,\bolQ)=-8J_1+12J_2 .
\end{align}
\end{subequations}
These interactions satisfy the conditions $\Gamma_1=\Gamma_2-|\Gamma_3|$ and $\Gamma_3>0$
for the range $-\frac{3}{2}\leq J_1/|J_2|\leq \frac{3}{2}$ and $J_2>0$.
Hence, for the large $S$ limit, the phase with infinite degeneracy
described by Eq.~(\ref{Inf:Deg1A}) appears.
The next-order correction in $1/S$ may remove this degeneracy.

The density of the condensed bosons near the saturation field is given by
\begin{equation}
\frac{\rho}{S}=\frac{\mu_s}{S\Gamma_1}=1-\frac{{\rm H}}{{\rm H}_{c1}}\ .
\end{equation}
The slop of the magnetization curve from the saturation field
can be understood from $M_s=-\VEV{S^z_l}/S=1-\rho/S$
and given by
\begin{equation}
\frac{dM_s}{d{\rm H}}=\frac{1}{{\rm H}_{c1}},
\end{equation}
which is directly connected to $M_s=0$ at ${\rm H}=0$, as shown in Fig.~\ref{Fig;magMF}.

\section{Formulation to treat a bound state}
\label{Ap:bm}
In this appendix, we give a technical review for the calculation of
the scattering amplitude $M(\Delta,\bolK;\bolp,\bolp^\prime)$ of two magnons.
This amplitude $M$ contains information about ground-state properties
of the single magnon BEC and also
information of two-magnon bound states;
a stable bound state is implied from a divergence of $M$,
whose residue represents the wave function of the bound state.\cite{BetheSalpeterReview}

\subsection{How to obtain $M$}
First, let us briefly review how to obtain
the scattering amplitude $M$.\cite{Batyev,Nikuni-Shiba-2,J1J2magBEC}
The bosonic Hamiltonian is given by Eq.~(\ref{Hboson}).
For $J_1/J_2\leq -3/2 $, $\epsilon_{min}=\epsilon(\bolQ_1)=4J_1+3J_2$ with
$\bolQ_1=(0,0,0)$ and,
for $J_1/J_2\geq -3/2 $, $\epsilon_{min}=\epsilon(\bolQ_2)=-3J_2$
with $\bolQ_2=(\pi,\pi,\pi)$.
The scattering amplitude is given by
\begin{equation}
\begin{split}
&M(\Delta,\bolK;\bolp,\bolp^\prime)=V(\bolp^\prime-\bolp)+V(-\bolp^\prime-\bolp)\\
&-\frac{1}{2}\int \frac{d^3 p^{\prime\prime}}{(4\pi)^3}
\frac{M(\Delta,\bolK;\bolp,\bolp^{\prime\prime})[V(\bolp^\prime-\bolp^{\prime\prime})
+V(-\bolp^\prime-\bolp^{\prime\prime})]}
{\omega(\bolK/2+\bolp^{\prime\prime})+\omega(\bolK/2-\bolp^{\prime\prime})+\Delta-i0^+},
\end{split}
\label{Ch2:laddereq}
\end{equation}
where the integral is taken for the region $p^{\prime\prime}_{x,y,z}\in (0,4\pi)$. This region laps the first Brillouin zone 4 times and the redundancy is accounted by $1/(4\pi)^3$.
In the following discussion, we abbreviate the arguments $\Delta$ and $\bolK$ in $M$ and denote $\frac{1}{N}\sum_{\bolp^\prime}$ as $\VEV{\ }$.

Taking the summation over $\bolp^\prime$ in Eq.~(\ref{Ch2:laddereq}), we obtain
\begin{equation}
\begin{split}
&\VEV{M}=4U\\
&\times(1-\frac{1}{2}\int \frac{d^3 p^{\prime\prime}}{(4\pi)^3}\frac{M(\bolp,\bolp^{\prime\prime})}
{\omega(\bolK/2+\bolp^{\prime\prime})+\omega(\bolK/2-\bolp^{\prime\prime})+\Delta-i0^+})\ ,
\end{split}
\end{equation}
where we have used $\VEV{\epsilon(\bolp^\prime)}=0$ and we have abbreviated $\bolp$ dependence of $\VEV{M}$.
Hence, taking the limit $U\rightarrow \infty$, we obtain
\begin{equation}
1-\frac{1}{2}\int \frac{d^3 p^{\prime\prime}}{(4\pi)^3}\frac{M(\bolp,\bolp^{\prime\prime})}
{\omega(\bolK/2+\bolp^{\prime\prime})+\omega(\bolK/2-\bolp^{\prime\prime})+\Delta-i0^+}=0.
\label{Ch2:UGammaeq1}
\end{equation}
Using this relation, we rewrite Eq.~(\ref{Ch2:laddereq}) as
\begin{equation}
\begin{split}
&M(\bolp,\bolp^\prime)=\VEV{M}+2\epsilon(\bolp^\prime-\bolp)+2\epsilon(-\bolp^\prime-\bolp)\\
&-\int \frac{d^3 p^{\prime\prime}}{(4\pi)^3}
\frac{M(\bolp,\bolp^{\prime\prime})[\epsilon(\bolp^\prime-\bolp^{\prime\prime})
+\epsilon(-\bolp^\prime-\bolp^{\prime\prime})]}
{\omega(\bolK/2+\bolp^{\prime\prime})+\omega(\bolK/2-\bolp^{\prime\prime})+\Delta-i0^+}.
\label{Ch2:UGammaeq2}
\end{split}
\end{equation}
Now, the problem is reduced to solving Eqs.~(\ref{Ch2:UGammaeq1})
and (\ref{Ch2:UGammaeq2}) simultaneously, which are free from the infinitely large $U$.

\begin{widetext}
Next, we expand $M(\bolp,\bolp^\prime)$ in the lattice harmonics.
Since
\begin{equation}
\begin{split}
&\epsilon(\bolp^\prime-\bolp)+\epsilon(-\bolp^\prime-\bolp)=2J_1 \left(\cos\frac{p_x+p_y+p_z}{2}\cos\frac{p^\prime_x+p^\prime_y+p^\prime_z}{2}
+\cos \frac{p_x+p_y-p_z}{2}\cos \frac{p^\prime_x+p^\prime_y-p^\prime_z}{2}\right.\\
&\left.+\cos \frac{p_x-p_y+p_z}{2}\cos \frac{p^\prime_x-p^\prime_y+p^\prime_z}{2}
+\cos \frac{-p_x+p_y+p_z}{2}\cos \frac{-p^\prime_x+p^\prime_y+p^\prime_z}{2}\right)\\
&+2J_2 (\cos p_x\cos p^\prime_x +\cos p_y\cos p_y^\prime +\cos p_z\cos p^\prime_z),
\end{split}
\end{equation}
we introduce
\begin{equation}
\begin{split}
M(\bolp,\bolp^\prime)=&\VEV{M}+A_1\cos \frac{p^\prime_x+p^\prime_y+p^\prime_z}{2}
+A_2\cos \frac{p^\prime_x+p^\prime_y-p^\prime_z}{2}
+A_3\cos \frac{p^\prime_x-p^\prime_y+p^\prime_z}{2}\\
&+A_4\cos \frac{-p^\prime_x+p^\prime_y+p^\prime_z}{2}
+A_5\cos p^\prime_x
+A_6\cos p^\prime_y
+A_7\cos p^\prime_z\ ,
\label{Ch2:Mexpand}
\end{split}
\end{equation}
where $\VEV{M}$ and $A_i$ are functions of $\Delta$, $\bolK$, and $\bolp$.
We also introduce
\begin{equation}
{\bf T}(\bolp)=(1,\ \cos \frac{p_x+p_y+p_z}{2},\ \cos \frac{p_x+p_y-p_z}{2},\
\cos \frac{p_x-p_y+p_z}{2},\ \cos \frac{-p_x+p_y+p_z}{2},\cos p_x,\cos p_y,\cos p_z)\ .
\end{equation}
Using this expression, Eq.~(\ref{Ch2:UGammaeq2}) reduces to
\begin{equation}
\begin{split}
&\sum_{i=2}^{8}\left(A_{i-1}+ \int \frac{d^3 p^{\prime\prime}}{(4\pi)^3}\frac{2J_{e_i}T_i(\bolp^{\prime\prime})}
{\omega(\bolK/2+\bolp^{\prime\prime})+\omega(\bolK/2-\bolp^{\prime\prime})+\Delta}
M(\bolp,\bolp^{\prime\prime})
-4J_{e_i}T_i(\bolp)\right)T_i(\bolp^\prime)=0\ ,
\label{gammaeq26}
\end{split}
\end{equation}
where $e_{2,3,4,5}=1$ and $e_{6,7,8}=2$.
To satisfy Eq.\ (\ref{gammaeq26}) for arbitrary $\bolp^\prime$,
the coefficients of the trigonometric functions of $\bolp^\prime$ must be $0$.
For convenience, we define
\begin{equation}
\tau_{ij}=\int \frac{d^3 p^{\prime\prime}}{(4\pi)^3} \frac{T_i(\bolp^{\prime\prime})T_j(\bolp^{\prime\prime})}
{\omega(\bolK/2+\bolp^{\prime\prime})+\omega(\bolK/2-\bolp^{\prime\prime})+\Delta}.
\end{equation}
Then, Eqs.~(\ref{Ch2:UGammaeq1}) and (\ref{Ch2:UGammaeq2}) are put together into
\begin{equation}
\left(
\begin{array}{cccccccc}
\tau_{11}/2 & \tau_{12}/2 & \tau_{13}/2 & \tau_{14}/2 & \tau_{15}/2 & \tau_{16}/2 & \tau_{17}/2 & \tau_{18}/2\\
2J_1\tau_{21} & 1+2J_1\tau_{22} & 2J_1\tau_{23} & 2J_1\tau_{24} & 2J_1\tau_{25}
& 2J_1\tau_{26} & 2J_1\tau_{27} & 2J_1\tau_{28}\\
2J_1\tau_{31} & 2J_1\tau_{32} & 1+2J_1\tau_{33} & 2J_1\tau_{34} & 2J_1\tau_{35} & 2J_1\tau_{36} & 2J_1\tau_{37} & 2J_1\tau_{38} \\
2J_1\tau_{41} & 2J_1\tau_{42} & 2J_1\tau_{43} & 1+2J_1\tau_{44} & 2J_1\tau_{45} & 2J_1\tau_{46} & 2J_1\tau_{47} & 2J_1\tau_{48}\\
2J_1\tau_{51} & 2J_1\tau_{52} & 2J_1\tau_{53} & 2J_1\tau_{54} & 1+2J_1\tau_{55}
& 2J_1\tau_{56} & 2J_1\tau_{57} & 2J_1\tau_{58}\\
2J_2\tau_{61} & 2J_2\tau_{62} & 2J_2\tau_{63} & 2J_2\tau_{64} & 2J_2\tau_{65}
& 1+2J_2\tau_{66} & 2J_2\tau_{67} & 2J_2\tau_{68}\\
2J_2\tau_{71} & 2J_2\tau_{72} & 2J_2\tau_{73} & 2J_2\tau_{74} & 2J_2\tau_{75}
& 2J_2\tau_{76} & 1+2J_2\tau_{77} & 2J_2\tau_{78}\\
2J_2\tau_{81} & 2J_2\tau_{82} & 2J_2\tau_{83} & 2J_2\tau_{84} & 2J_2\tau_{85}
& 2J_2\tau_{86} & 2J_2\tau_{87} & 1+2J_2\tau_{88}\\
\end{array}
\right)
\left(
\begin{array}{c}
\VEV{M}\\
A_1\\
A_2\\
A_3\\
A_4\\
A_5\\
A_6\\
A_7\\
\end{array}
\right)
=
\left(
\begin{array}{c}
1\\
4T_2(\bolp)\\
4T_3(\bolp)\\
4T_4(\bolp)\\
4T_5(\bolp)\\
4T_6(\bolp)\\
4T_7(\bolp)\\
4T_8(\bolp)\\
\end{array}
\right).
\label{Ch2:Mmatrix1}
\end{equation}

In the case $\bolK=\bolK_0=(2\pi,2\pi,2\pi)$, we can simplify the above calculation.
Since
$\omega(\bolK_0/2+\bolp^{\prime\prime})+\omega(\bolK_0/2-\bolp^{\prime\prime})
=-2J_2(\cos p_x^{\prime\prime} +\cos p_y^{\prime\prime}+\cos p_z^{\prime\prime})-2\mu
$, at $\bolp=0$ we use
\begin{equation}
\begin{split}
M(\bolp=0,\bolp^\prime)=&\VEV{M}+4A_1\cos \frac{p_x^\prime}{2}\cos \frac{p_y^\prime}{2}\cos \frac{p_z^\prime}{2}+A_2(\cos p^\prime_x
+\cos p^\prime_y
+\cos p^\prime_z)\ .
\end{split}
\end{equation}
Then, we obtain
\begin{equation}
\begin{split}
&\left[A_1+ \int \frac{d^3 p^{\prime\prime}}{(4\pi)^3}
\frac{\frac{J_1}{2}T^\prime_2(\bolp^{\prime\prime})}
{\omega(\bolK_0/2+\bolp^{\prime\prime})+\omega(\bolK_0/2-\bolp^{\prime\prime})+\Delta}
M(0,\bolp^{\prime\prime})-4J_1\right] T^\prime_2(\bolp^\prime)\\
+&\left[ A_2+\int \frac{d^3 p^{\prime\prime}}{(4\pi)^3}
\frac{\frac{2J_2}{3}T_3^\prime(\bolp^{\prime\prime})}
{\omega(\bolK_0/2+\bolp^{\prime\prime})+\omega(\bolK_0/2-\bolp^{\prime\prime})+\Delta}
M(0,\bolp^{\prime\prime})-4J_2\right] T^\prime_3(\bolp^{\prime})=0,
\end{split}
\end{equation}
where
\begin{equation}
{\bf T}^\prime(\bolp)=
(1,\ 4\cos \frac{p_x}{2}\cos \frac{p_y}{2}\cos \frac{p_z}{2},\ \cos p_x+\cos p_y+\cos p_z).
\end{equation}
\end{widetext}
Eventually, $M$ is understood by solving
\begin{align}
L
\left(
\begin{array}{c}
\VEV{M}\\
A_1\\
A_2
\end{array}
\right)
=
\left(
\begin{array}{c}
1\\
4J_1\\
4J_2
\end{array}
\right)
\label{Mvec}
\end{align}
with
\begin{align}
L&=
\left(
\begin{array}{ccc}
\tau^\prime_{11}/2 & \tau^\prime_{12}/2 & \tau^\prime_{13}/2 \\
J_1\tau^\prime_{21}/2 & 1+J_1\tau^\prime_{22}/2 & J_1\tau^\prime_{23}/2 \\
2J_1\tau^\prime_{31}/3 & 2J_1\tau^\prime_{32}/3 & 1+2J_1\tau^\prime_{33}/3
\end{array}
\right),\\
\tau_{ij}^\prime&=\int \frac{d^3 p^{\prime\prime}}{(4\pi)^3} \frac{T^\prime_i(\bolp^{\prime\prime})T^\prime_j(\bolp^{\prime\prime})}
{\omega(\bolK_0/2+\bolp^{\prime\prime})+\omega(\bolK_0/2-\bolp^{\prime\prime})+\Delta}\ .
\end{align}
Due to the symmetry, we obtain $\tau^\prime_{12}=\tau^\prime_{23}=\tau^\prime_{21}=\tau^\prime_{32}=0$.
Hence, the divergence of $M$ (or $A_1$) appears when $L_{22}$ vanishes, i.e., $L_{22}=0$.
As a result, the binding energy of two magnons is given by solving
$L_{22}=1+J_1\tau^\prime_{22}/2=0$, i.e.,
\begin{equation}
\begin{split}
J_1\int \frac{d^3 p}{\pi^3}\frac{\cos ^2\frac{p_x}{2}\cos^2\frac{p_y}{2}\cos^2\frac{p_z}{2}}
{\omega(\bolK_0/2+\bolp)+\omega(\bolK_0/2-\bolp)-\Delta}=1.
\end{split}
\label{eq:b_energy}
\end{equation}

We note that a bound state could be also stable if
\begin{align}
\text{det}\left(
\begin{array}{cc}
L_{11} & L_{13}\\
L_{31} & L_{33}
\end{array}
\right)
=0 \ .
\label{Ap:L13}
\end{align}
However, we numerically confirmed that Eq.~(\ref{Ap:L13})
is never satisfied for any $J_1/J_2$.
Thus Eq.~(\ref{eq:b_energy}) is the only condition for the existence of
two-magnon bound states.

\subsection{Wave function of bound state}
Next, let us discuss the wavefunction of the bound magnon state.
Since there is a detailed-review paper\cite{BetheSalpeterReview}
of the formulation to treat a bound state from the Bethe-Salpeter equation,
we only concentrate on the technical point related to
bound magnons on the fully saturated ferromagnetic phase.

\begin{figure}[tb]
\begin{center}
\includegraphics[scale=0.3]{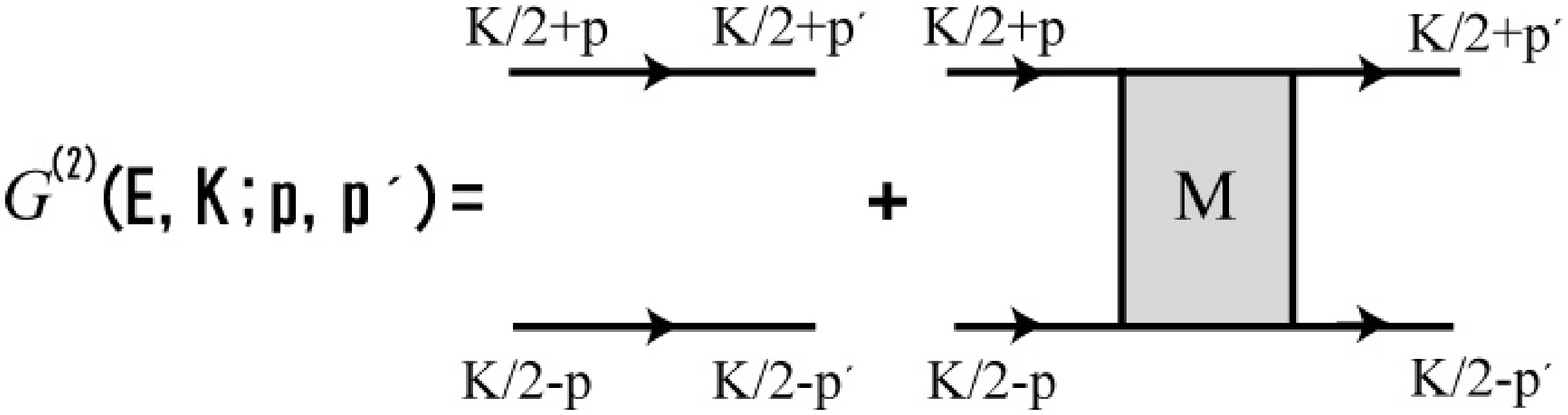}
\caption{Two-magnon Green's function on the fully polarized ferromagnetic state.
\label{Greentwo}
}
\end{center}
\end{figure}

The free two-body Green's function in the ferromagnetic state of
the energy $E$ and the center-of-mass momentum $\bolK$ is written as
\begin{equation}
\begin{split}
iG^{(2)}_0(E,\bolK;\bolp,\bolp^\prime)
=iG^{(2)}_0(E,\bolK;\bolp)\frac{\delta(\bolp-\bolp^\prime)+\delta(\bolp+\bolp^\prime)}{2},
\end{split}
\end{equation}
where
\begin{equation}
\begin{split}
iG^{(2)}_0(E,\bolK;\bolp)=\frac{2i}{E-[\omega(\bolK+\bolp)+\omega(\bolK-\bolp)-2\mu]+i0^+}.
\end{split}
\end{equation}
In the interacting case, the two-body Green's function shown in Fig.~\ref{Greentwo} reads
\begin{align*}
&iG^{(2)}(E,\bolK;\bolp,\bolp^\prime)=iG^{(2)}_0(E,\bolK;\bolp,\bolp^\prime)\\
&+\frac{1}{4}iG^{(2)}_0(E,\bolK;\bolp,\bolp^{\prime\prime}) [-iM(\Delta=-E-2\mu,\bolK;\bolp^{\prime\prime},\bolp^{\prime})] \\
&\ \ \ \ \ \times iG^{(2)}_0(E,\bolK;\bolp^{\prime\prime},\bolp^\prime)\ ,
\end{align*}
where the summation over the repeated momentum is implied.
In what follows, we may abbreviate the arguments in $M$, $G^{(2)}$,
and $G_0^{(2)}$ for convenience.
The divergence of $M(\Delta=\Delta_B,\bolK)$ leads to a divergence
of the Green's function that implies the existence of a stable two-magnon bound state
with the dispersion $\Delta_B(\bolK)$. The wavefunction $\chi_\bolK(\bolp)$ of
the bound state is understood from the residue:\cite{BetheSalpeterReview}
\begin{equation}
\frac{1}{4}G^{(2)}_0 M G^{(2)}_0 \Big|_{\Delta\rightarrow \Delta_B}=\frac{\chi_\bolK(\bolp^\prime){\chi_\bolK}^\dagger(\bolp)}{\Delta-\Delta_B}\ ,
\end{equation}
where we assumed the limit $\Delta \rightarrow \Delta_B$.
Considering the wavefunction $\chi_{\bolK}(\bolp)$ of the bound state
at $\bolK=\bolK_0=(2\pi, 2\pi, 2\pi)$, we obtain
\begin{equation}
\begin{split}
&\chi_{\bolK_0}(\bolp^\prime){\chi_{\bolK_0}}^\dagger(\bolp=0)=\frac{\Delta-\Delta_B}{4}G^{(2)}_0 M G^{(2)}_0\Big|_{\Delta\rightarrow \Delta_B}\\
&=\frac{1}{4}
G_0^{(2)}(\bolp^\prime)
\left[\bolT^\prime (\bolp^\prime)
\{(\Delta-\Delta_B)L^{-1}\}\Big|_{\Delta\rightarrow \Delta_B}
\left(
\begin{array}{c}
1\\
4J_1 \\
4J_2 \\
\end{array}
\right) \right]\\
&\ \ \ \ \ \times G_0^{(2)}(0),
\end{split}
\label{Ch2:FormG0}
\end{equation}
where we have used Eq.~(\ref{Mvec}).
Since the divergence of $M$ ($L^{-1}$) occurs due to
$L_{22}=O(\Delta-\Delta_B)\rightarrow 0$ and $L_{2i}=L_{i2}=0$ for $i={1,3}$,
$(\Delta-\Delta_B)L^{-1}$ is given by
\begin{equation}
(\Delta-\Delta_B)L^{-1}\Big|_{\Delta\rightarrow \Delta_B}=\left(
\begin{array}{ccc}
0 & 0 & 0\\
0 & c_1 & 0\\
0 & 0 & 0\\
\end{array}
\right),
\end{equation}
where $c_1$ is a numerical constant, and the matrix elements of (1,1), (1,3), (3,1), (3,3)
vanish in the limit $(\Delta-\Delta_B)\rightarrow 0$.
Hence, the wavefunction is given by
\begin{equation}
\begin{split}
\chi_{\bolK_0}(\bolp)\propto
G_0^{(2)}\textbf{(}-2\mu-\Delta_B(\bolK_0),\bolK_0;\bolp\textbf{)}\cos \frac{p_x}{2}\cos \frac{p_y}{2}\cos \frac{p_z}{2}.
\end{split}
\end{equation}



\begin{thebibliography}{101}
\bibitem{footnote2}
We use the terminology ``single magnon BEC" for the BEC induced by
the magnon that have $\VEV{S^z}=1$.
When the instability at the saturation field is governed by single magnons,
condensation of magnons below the saturation field corresponds to the single-magnon BEC.
%
In the quantum systems,
the single-magnon BEC leads to a canted antiferromagnetic order.
%
As a concrete example, see Sec.~\ref{Sec:singleBEC}.
\end{thebibliography}
\end{document}